\documentclass[prl,twocolumn,showpacs,floatfix,superscriptaddress]{revtex4}

\usepackage{graphicx}
\def\mpl{\ifmmode \overline M_{Pl}\else $\overline M_{Pl}$\fi}

\newcommand{\pom}{I\!\! P}

\begin{document}
\bibliographystyle{revtex}
\input{psfig.sty}
\title{Central pseudorapidity gaps in events with a leading antiproton 
at the Fermilab Tevatron $\bar pp$ Collider\\
\vglue 0.15in
The CDF Collaboration\\
\vglue 0.15in
(Submitted to Physical Review Letters)}

\font\eightit=cmti8
\def\r#1{\ignorespaces $^{#1}$}
\hfilneg
\author{
\noindent
D.~Acosta,\r {14} T.~Affolder,\r {25} H.~Akimoto,\r {50}
M.~G.~Albrow,\r {13} D.~Ambrose,\r {37}   
D.~Amidei,\r {28} K.~Anikeev,\r {27} J.~Antos,\r 1 
G.~Apollinari,\r {13} T.~Arisawa,\r {50}
% ...
}
\affiliation{\r 1  {\eightit Institute of Physics, Academia Sinica, 
Taipei, Taiwan 11529, Republic of China} \\
\r 2  {\eightit Argonne National Laboratory, Argonne, Illinois 60439} \\
\r 3  {\eightit Istituto Nazionale di Fisica Nucleare, University of 
Bologna, I-40127 Bologna, Italy} \\
\r 4  {\eightit Brandeis University, Waltham, Massachusetts 02254} \\
\r 5  {\eightit University of California at Davis, Davis, California  95616} \\
\r 6  {\eightit University of California at Los Angeles, Los 
Angeles, California  90024} \\ 
\r 7  {\eightit University of California at Santa Barbara, 
Santa Barbara, California 
93106} \\ 
\r 8 {\eightit Instituto de Fisica de Cantabria, CSIC-University of Cantabria, 
39005 Santander, Spain} \\
\r 9  {\eightit Carnegie Mellon University, Pittsburgh, Pennsylvania  15213} \\
\r {10} {\eightit Enrico Fermi Institute, University of Chicago, Chicago, 
Illinois 60637} \\
\r {11}  {\eightit Joint Institute for Nuclear Research, RU-141980 Dubna, Russia}
\\
\r {12} {\eightit Duke University, Durham, North Carolina  27708} \\
\r {13} {\eightit Fermi National Accelerator Laboratory, Batavia, Illinois 
60510} \\
\r {14} {\eightit University of Florida, Gainesville, Florida  32611} \\
\r {15} {\eightit Laboratori Nazionali di Frascati, Istituto Nazionale di Fisica
               Nucleare, I-00044 Frascati, Italy} \\
\r {16} {\eightit University of Geneva, CH-1211 Geneva 4, Switzerland} \\
\r {17} {\eightit Glasgow University, Glasgow G12 8QQ, United Kingdom}\\
\r {18} {\eightit Harvard University, Cambridge, Massachusetts 02138} \\
\r {19} {\eightit Hiroshima University, Higashi-Hiroshima 724, Japan} \\
\r {20} {\eightit University of Illinois, Urbana, Illinois 61801} \\
\r {21} {\eightit The Johns Hopkins University, Baltimore, Maryland 21218} \\
\r {22} {\eightit Institut f\"{u}r Experimentelle Kernphysik, 
Universit\"{a}t Karlsruhe, 76128 Karlsruhe, Germany} \\
\r {23} {\eightit High Energy Accelerator Research Organization (KEK), Tsukuba, 
Ibaraki 305, Japan} \\
\r {24} {\eightit Center for High Energy Physics: Kyungpook National
University, Taegu 702-701; Seoul National University, Seoul 151-742; and
SungKyunKwan University, Suwon 440-746; Korea} \\
\r {25} {\eightit Ernest Orlando Lawrence Berkeley National Laboratory, 
Berkeley, California 94720} \\
\r {26} {\eightit University College London, London WC1E 6BT, United Kingdom} \\
\r {27} {\eightit Massachusetts Institute of Technology, Cambridge,
Massachusetts  02139} \\   
\r {28} {\eightit University of Michigan, Ann Arbor, Michigan 48109} \\
\r {29} {\eightit Michigan State University, East Lansing, Michigan  48824} \\
\r {30} {\eightit Institution for Theoretical and Experimental Physics, ITEP,
Moscow 117259, Russia} \\
\r {31} {\eightit University of New Mexico, Albuquerque, New Mexico 87131} \\
\r {32} {\eightit Northwestern University, Evanston, Illinois  60208} \\
\r {33} {\eightit The Ohio State University, Columbus, Ohio  43210} \\
\r {34} {\eightit Osaka City University, Osaka 588, Japan} \\
\r {35} {\eightit University of Oxford, Oxford OX1 3RH, United Kingdom} \\
\r {36} {\eightit Universita di Padova, Istituto Nazionale di Fisica 
          Nucleare, Sezione di Padova, I-35131 Padova, Italy} \\
\r {37} {\eightit University of Pennsylvania, Philadelphia, 
        Pennsylvania 19104} \\   
\r {38} {\eightit Istituto Nazionale di Fisica Nucleare, University and Scuola
               Normale Superiore of Pisa, I-56100 Pisa, Italy} \\
\r {39} {\eightit University of Pittsburgh, Pittsburgh, Pennsylvania 15260} \\
\r {40} {\eightit Purdue University, West Lafayette, Indiana 47907} \\
\r {41} {\eightit University of Rochester, Rochester, New York 14627} \\
\r {42} {\eightit Rockefeller University, New York, New York 10021} \\
\r {43} {\eightit Instituto Nazionale de Fisica Nucleare, Sezione di Roma,
University di Roma I, ``La Sapienza," I-00185 Roma, Italy}\\
\r {44} {\eightit Rutgers University, Piscataway, New Jersey 08855} \\
\r {45} {\eightit Texas A\&M University, College Station, Texas 77843} \\
\r {46} {\eightit Texas Tech University, Lubbock, Texas 79409} \\
\r {47} {\eightit Istituto Nazionale di Fisica Nucleare, 
University of Trieste, Udine, Italy} \\
\r {48} {\eightit University of Tsukuba, Tsukuba, Ibaraki 305, Japan} \\
\r {49} {\eightit Tufts University, Medford, Massachusetts 02155} \\
\r {50} {\eightit Waseda University, Tokyo 169, Japan} \\
\r {51} {\eightit University of Wisconsin, Madison, Wisconsin 53706} \\
\r {52} {\eightit Yale University, New Haven, Connecticut 06520} \\
}
\begin{abstract}
We report a measurement 
of the fraction of events with a large
pseudorapidity gap $\Delta \eta$ within the pseudorapidity 
region available to the proton 
dissociation products $X$ in $\bar p+p\rightarrow \bar p+X$.
For a final state $\bar p$ of fractional 
momentum loss $\xi_{\bar p}$ and 4-momentum transfer 
squared $t_{\bar p}$ within 
$0.06<\xi_{\bar p}<0.09$ and $|t_{\bar p}|<1.0$ [0.2] GeV$^2$ at 
$\sqrt s=1800$ [630] GeV,
the fraction of events with $\Delta \eta>3$
is found to be $0.246\pm 0.001\,{\rm (stat)}\pm 0.042\,{\rm (syst)}$ 
[$0.184\pm 0.001\,{\rm (stat)}\pm 0.043\,{\rm (syst)}$].
Our results are compared with gap
fractions measured in minimum bias $\bar pp$ collisions 
and with theoretical expectations.

\end{abstract}


\maketitle

\date{today}

In a previous Letter~\cite{CDF_DD}, 
we reported a measurement of the fraction of 
events with a central pseudorapidity gap 
$\Delta\eta$~\cite{pseudo} produced in 
$\bar pp$ collisions at $\sqrt s=1800$ and 630 GeV. Here, we present 
results from a similar measurement performed in a sub-sample of $\bar pp$
events containing a leading (high longitudinal momentum) antiproton (Fig.~1). 
Large pseudorapidity gaps are presumed to be due to Pomeron 
($\pom$) exchange and are the signature for diffraction~\cite{Regge}.
The process with a leading beam particle in the final state, 
which is kinematically associated  with an adjacent pseudorapidity gap, 
is known as single diffraction dissociation (SD), while that with a 
central gap as double diffraction dissociation (DD). The process in Fig.~1 is 
a combination of $\bar p$-$p$ SD and $\pom$-$p$ DD 
and will be referred to in this paper as SDD.

\begin{figure}
\centerline{\psfig{figure=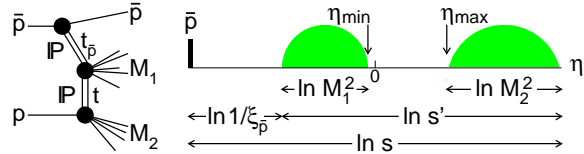,width=3in}}
\vglue -2in
\caption{Schematic diagram and event topology in pseudorapidity space of a
SDD (single diffraction plus gap) interaction,
$\bar p+p\rightarrow \bar p+{\rm GAP}_{\bar p}+M_1+{\rm GAP}+M_2$, 
with a leading outgoing 
antiproton of fractional momentum loss $\xi_{\bar p}$, associated with a 
pseudorapidity gap 
$\Delta\eta_{\bar p}=\ln\frac{1}{\xi_{\bar p}}$, and a gap 
within the region of $\eta$ spanned by 
$\ln{s'}=\ln s-\ln\frac{1}{\xi_{\bar p}}$.}
\end{figure}
Low transverse momentum ($p_T$) processes have traditionally been treated 
theoretically in the framework of Regge theory~\cite{Regge}.  
The introduction of a linear Pomeron ($\pom$) 
trajectory, $\alpha(t)=\alpha(0)+\alpha't$,
with intercept $\alpha(0)=1+\epsilon>1$,
enables the theory to correctly predict
certain salient features of the high energy behavior of hadronic 
interactions, such as the rise of the total cross section  
and the shrinking of the forward elastic scattering peak 
with increasing energy.  However, the success of the theory in 
describing diffraction has been limited. While the shape of the SD and DD 
distributions as a function of $\Delta\eta$
are correctly described, the normalization was found to be suppressed 
relative to the theoretical predictions by about an order of magnitude 
as the energy increases from $\sqrt s \sim 20$ to 
1800 GeV~\cite{CDF_DD,CDF_PRD,R}.
Proposals made to address this issue are divided into two general groups:
(a) those based on ``damping" of the SD cross section at small 
(anti)proton fractional momentum loss
$\xi$ or on a changing Pomeron intercept as a function of 
$\sqrt s$~\cite{ES} or as a function of $\xi$~\cite{Tan}, 
and (b) those in which only the 
overall normalization is suppressed as $\sqrt s$  
increases~\cite{R,K,GLM,multigap}. 
The models of group (a) cannot be applied to SDD.
In a parton model approach developed by Bjorken~\cite{Bj} 
to describe events with a large pseudorapidity gap between two jets
produced by a colorless two-gluon exchange,  
a suppression of the overall normalization relative to 
the QCD calculation was predicted due to 
additional partonic {\em color} exchanges in the same event which 
spoil the diffractive pseudorapidity gap signature.
In this model, which belongs to group (b),
the color exchanges would simultaneously 
spoil all diffractive rapidity
gaps in an event and therefore the ratio of two-gap to one-gap events 
in a given interaction would be unaffected. 
This approach should also hold for low $p_T$ diffractive processes. 
In this Letter, we examine this issue by 
studying $\bar p+p\rightarrow \bar p+X$ with/without a gap 
$\Delta\eta$ within the $\eta$ range of the system $X$ 
in addition to the gap of nominal~\cite{pseudo} value  
$\Delta\eta_{\bar p}= -\ln{\xi_{\bar p}}$ expected to be 
associated with the leading final state antiproton.

Our study is based on our $\sqrt s=$1800 (630) GeV inclusive SD data
described in Ref. \cite{CDF_JJ_RP_1800} (\cite{CDF_JJ_RP_630}).
The events were collected in the 1995-96 Tevatron Run 1C
by triggering the Collider Detector at Fermilab (CDF) on an antiproton
detected in a Roman Pot Spectrometer (RPS)~\cite{CDF_JJ_RP_1800}.
The average instantaneous luminosity during event collection was
$0.2\times 10^{30}\;(1.5\times 10^{30})$ cm$^{-2}$ sec$^{-1}$
at $\sqrt s=$1800 (630) GeV.
The components of CDF~\cite{detector} relevant to this study
are the central tracking chamber (CTC), the calorimeters,
and two scintillation beam-beam counter (BBC)
arrays.
The CTC tracking efficiency varies from $\sim60$\% for $p_T=300$
MeV to over 95\% for $p_T>400$ MeV within $|\eta|<1.2$, and
falls monotonically beyond
$|\eta|=1.2$ approaching zero at $|\eta|\sim 1.8$.
The calorimeters have projective tower geometry and cover the
regions $|\eta|<1.1$ (central), $1.1<|\eta|<2.4$ (plug), and $2.2<|\eta|<4.2$
(forward). The $\Delta \eta \times \Delta \phi$ tower dimensions, 
where $\phi$ is the azimuthal angle,  are approximately
$0.1\times 15^{\circ}$ for the central and $0.1\times 5^{\circ}$ for the
plug and forward calorimeters. The BBC arrays cover 
the region $3.2<|\eta|<5.9$.

The events were required to have a reconstructed RPS track of 
$0.06<\xi_{\bar p}<0.09$ and $|t_{\bar p}|<1.0$ [0.2] GeV$^2$ at 
$\sqrt s=1800$ [630] GeV, a hit on $BBC_p$ (proton-side BBC) to exclude 
DPE events, and no more than one reconstructed vertex 
within $\pm 60$ cm from the center of the detector along the beam direction. 
The vertex requirement 
was imposed to reject {\em overlap}
events due to multiple interactions in the same beam-beam crossing, since 
additional interactions would most likely 
spoil the pseudorapidity gap signature of a diffractive event.

At $\sqrt s=$1800 (630) GeV, the average $\xi_{\bar p}$ value of 0.075 
of our data samples 
corresponds to $\pom$-$p$ collision energies of  
$\sqrt{s'}=\sqrt{\xi_{\bar p} s}=493$ (173) GeV,
at which the proton dissociation  
products cover the nominal $\eta$-range from the maximum of  
$\eta=\ln \sqrt s=7.5$ (6.5) down to $\eta= -\ln(\xi\sqrt s)=-4.9$ (-3.9).    
Thus, the CDF calorimeter coverage, $|\eta|<4.2$, is well suited   
for the present study.

Our analysis is similar to that used in evaluating the DD fraction in 
minimum bias events collected with a BBC coincidence trigger~\cite{CDF_DD}.
The method we use is based on the approximately
flat dependence of the event rate on $\Delta\eta$ expected for SDD events
compared to the exponential dependence expected for
the normal SD events where rapidity gaps within the diffractive cluster $X$ 
are due to random multiplicity fluctuations. 
Thus, in a plot of event rate versus $\Delta\eta$, the SDD signal
will appear as a flattening of an exponentially
falling distribution at large $\Delta\eta$~\cite{Bj}.
In order to independently monitor detector effects 
in the positive and negative $\eta$ directions,  
we look for $\eta_{max}$ ($\eta_{min}$), the $\eta$ of the 
particle closest to $\eta=0$ in the proton (antiproton) direction,
and measure {\em experimental} gaps overlapping $\eta=0$, 
$\Delta\eta_{exp}^0\equiv\eta_{max}-\eta_{min}$ (see Fig.~1).
For this purpose, a particle is defined as a reconstructed
track in the CTC, a calorimeter tower with energy above a given threshold,
or a BBC hit. The tower energy thresholds used, chosen to lie
comfortably above noise level, 
are $E_T=0.2$ GeV for the central and plug and
$E=1$ GeV for the forward calorimeters. The calorimeter noise was measured 
using beam-beam crossing events with no reconstructed vertex. 
At the calorimeter interfaces near
$|\eta|\sim 0$, 1.1  and $\sim 2.4$, where the noise level was found to 
be higher, we use higher thresholds of up to 0.3 GeV. 
The average number of calorimeter towers per unit $\Delta\eta$ with 
$E_T^{noise}$ above threshold in an event is 
$\sim 0.07$, small compared to the corresponding 
average particle density of $\sim 3$ in the data.
The fraction of SDD to total number of events based on $\Delta\eta_{exp}^0$ 
is obtained directly from the data and 
corrected for (a) contamination from SD events, (b) the effect 
of the unobserved (below threshold) particles, 
and (c) the triggering efficiency (acceptance) of $BBC_p$
for SD and SDD events. These corrections are made using a hadron-level 
MC simulation.

The MC generation of SD events is described in Ref.~\cite{CDF_PRD}. 
For SDD events, a diffractive $\bar p$
and a cluster of $M^2_X=s\xi$ are generated as for a SD interaction, 
and a DD interaction is assumed to take place in the $\pom$-$p$ collision,
which is treated as in Ref.~\cite{CDF_DD} and boosted to the lab frame. 
The same thresholds are used for particles in the MC as 
for towers in the data, after dividing the 
generated particle $E_T$ by an $\eta$-dependent 
calibration coefficient of average value $\sim 1.6$ 
representing the ratio of true to measured
calorimeter energy. 
The MC generator includes the calorimeter noise,  
and for charged particles it is followed by a detector simulation.

Figure~2 shows lego histograms of events versus $\eta_{max}$
and $-\eta_{min}$ for (a) data and (b) MC generated events,
as well as  MC events for (c) only SD and (d) only SDD  
at $\sqrt{s}=1800$ GeV. Similar results are obtained at $\sqrt s=630$ GeV.
The observed structure in the
distributions along $\eta_{max(min)}$ is caused by the variation of the tower
energy thresholds with $|\eta|$. The bins at $|\eta_{max(min)}|=3.3$ contain
all events within the BBC range of $3.2<|\eta_{max(min)}|<5.9$.

\begin{figure}[htb]
\centerline{\psfig{figure=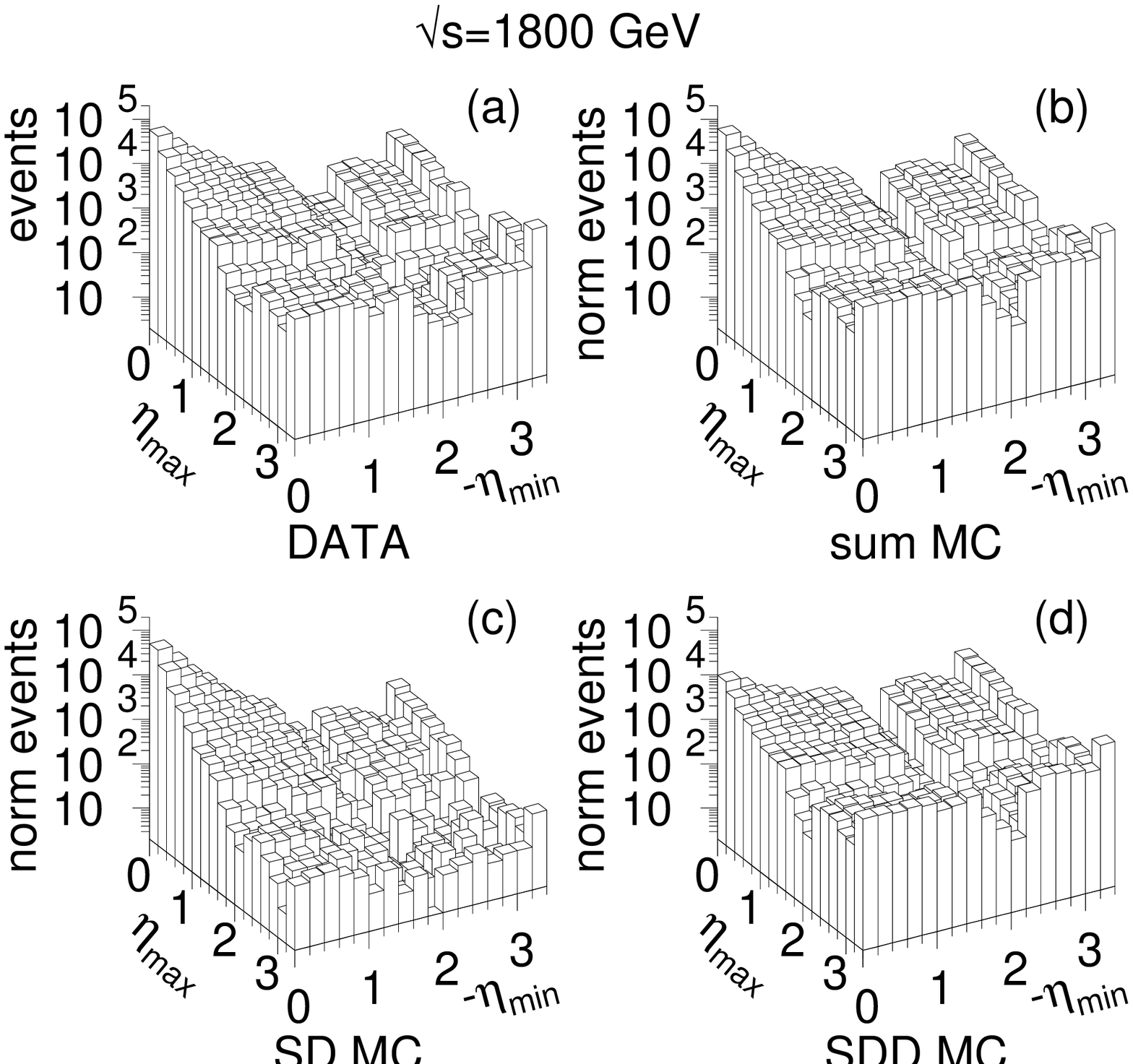,width=3in}}
\caption{The number of events as a function of \protect$\eta_{max}$ and
\protect$-\eta_{min}$,
the $\eta$ of the track or hit tower closest to $\eta=0$ in the proton
and antiproton direction, respectively, at $\protect\sqrt s=1800$ GeV:
(a) data; (b) Monte Carlo simulation; (c,d) the individual contributions of 
Monte Carlo generated SD and SDD (SD + gap) events. 
The MC distributions are normalized by a two-component fit to the data using 
distributions (c) and (d).}
\end{figure}
Figure~3 presents the number of events
as a function of $\Delta\eta^0_{exp}$ for
the 1800 GeV data (points) and for a fit to the data using
a mixture of MC generated SD and SDD 
contributions (solid histogram). 
The SD contribution (dashed histogram) exhibits
an approximately exponential fall with 
increasing $\Delta\eta^0_{exp}$, as expected.
The region of $\Delta\eta^0_{exp}>3$ is dominated by the SDD signal
and is used to extract the gap fraction (ratio of SDD to total number 
of events).
The approximately flat behaviour expected for the SDD distribution 
in this region 
is modulated by the $\eta$-dependent tower thresholds used, causing the 
observed bumps and dips, and by the $BBC_p$ acceptance for SDD events, 
which decreases with increasing $\Delta\eta^0_{exp}$. 
The MC simulation reproduces  
these features of the data.

\begin{figure}[htb]
\centerline{\psfig{figure=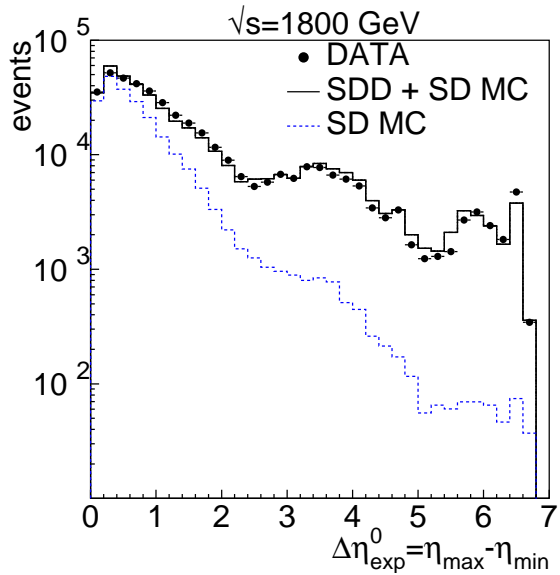,width=3in}}
\caption{The number of events as a function of
$\Delta\eta_{exp}^0=\eta_{max}-\eta_{min}$ for data at
$\protect\sqrt s=1800$ GeV (points), for SDD (SD + gap)  MC generated events 
(solid line), and for only SD MC events (dashed line).}
\end{figure}
At $\sqrt s=1800$ [630] GeV, the fraction of events with
$\Delta\eta^0_{exp}>3$ 
is $(15.9\pm 0.1)\%$ [$(17.5\pm 0.2)\%$], of which
the contribution of background
SD events, estimated using the MC simulation,
is $1.2\%$ [$2.4\%$].
The quoted errors are statistical.
The amount of SD background in the region $\Delta\eta^0_{exp}>3$
depends on the tower energy calibration coefficients and thereby on the
calorimeter tower energy thresholds used in the MC.
For example, increasing these thresholds has the effect of decreasing
the multiplicity in the MC generated events, resulting in larger rapidity gaps
and hence larger SD background in the region of $\Delta\eta^0_{exp}>3$.
The systematic uncertainty in the background
is estimated by raising (lowering) the tower thresholds in the MC
by a factor of 1.25 (evaluated from a multiplicity uncertainty of 
$\pm 10\%$), which increases (decreases) the background by a factor of 1.6.

The correction factors needed to account for the effect of unobserved
particles and thus convert the measured gap fractions
to gap fractions corresponding to the nominal gap definition~\cite{pseudo},
$\Delta\eta^0\equiv \ln\frac{s's_{\circ}}{M_1^2M_2^2}$
($\ln\frac{M_i^2}{\sqrt {s's_{\circ}}}<0,\; i=1,2$), where $s_0=1$ GeV$^2$, 
were evaluated using the SDD Monte Carlo simulation and
found to be $0.81$ [$0.73$] 
at $\sqrt s=1800$ [630] GeV. Systematic errors are obtained by 
varying the tower energy thresholds by $\pm 25\%$. These errors are 
correlated with the errors in the SD background contamination and therefore 
a combined systematic uncertaity of 15\% [23\%] was evaluated for both 
these effects and 
applied to the extracted nominal gap fractions. 

The $BBC_p$ acceptance, evaluated from the SDD MC simulation,
is $0.68\pm 0.06({\rm syst})$ [$0.81\pm 0.04({\rm syst})$], where
the error is due to a 20\% systematic uncertainty assigned to the 
fraction of the diffraction dissociation mass clusters 
which do not trigger the 
$BBC_p$. For SD events, the BBC acceptance is 
$0.98\pm 0.01({\rm syst})$ [$0.98\pm 0.01({\rm syst})$].
Including all systematic errors, the acceptance-corrected SDD fractions for
nominal gaps $\Delta\eta^0>3$ are 
$0.174\pm 0.001\,{\rm (stat)}\pm 0.030\,{\rm (syst)}$ 
[$0.138\pm 0.001\,{\rm (stat)}\pm 0.032\,{\rm (syst)}$] at $\sqrt s=1800$ 
[630] GeV. 

The $\Delta\eta^0>3$ fractions are extrapolated to all SDD gaps 
of $\Delta\eta>3$ using the shape of the gap distribution of Eq.~1, 
which is based on Regge theory and factorization. 
This equation, which was used in the MC simulation, 
is obtained from the equation for SD by replacing the  
$\pom$-$p$ total cross section factor 
with the $\pom$-$p$ DD factor $\kappa \{...\}$ (see Ref.~\cite{CDF_DD}). 
\begin{eqnarray}
\frac{d^5\sigma}{dt_{\bar p}dtd\xi_{\bar p} d\Delta\eta d\eta_c}=
\left[{\beta(t)\over 4\sqrt{\pi}}\, e^{[\alpha(t_{\bar p})-1]
\ln\frac{1}{\xi}}\right]^2\times \nonumber\\
\kappa \left\{
\kappa\left[{\beta(0)\over 4\sqrt{\pi}} e^{[\alpha(t)-1]
\Delta\eta}\right]^2
\;\kappa\left[\beta^{2}(0){{\left(\frac{\textstyle{s''}}
{\textstyle s_{\circ}}\right)}}^{\epsilon}\right]\right\}
\end{eqnarray}
Here, $\eta_c$ is the center of the 
gap $\Delta\eta$, $\beta(0)$ the $\pom$-$p$ coupling, $\kappa$ the ratio 
of the triple-Pomeron to the $\pom$-$p$ couplings, 
and $\ln\frac{s''}{s_0}=\ln\frac{s}{s_0}-
\ln\frac{1}{\xi_{\bar p}}-\Delta\eta$ the rapidity space 
occupied by particles; $\sqrt{s''}$ will be referred to below as 
``diffractive sub-energy''.
For numerical evaluations we use~\cite{CDF_DD}
$\epsilon=0.104\pm 0.002$, 
$\alpha'=0.25$ GeV$^{-2}$, $\kappa=0.17$, $\beta(t_{\bar p})=
6.57\;{\rm GeV}^{-1}\;[4.1\;{\rm mb}^{\frac12}]\times F_1(t_{\bar p})$, 
where $F_1(t_{\bar p})$ is the nucleon form factor. 
The variable $t$ is not measured and therefore 
is integrated over in the calculations. 
Owing to the increased phase space resulting from 
releasing the requirement that 
the rapidity gap overlap $\eta=0$, the $\Delta\eta>3$ fractions are 
found to be larger than the $\Delta\eta^0>3$ ones by a factor of 1.44 [1.40] 
at $\sqrt s=1800$ [630] GeV. The evaluation of this 
factor is performed analytically and therefore no 
error is assigned to it; the effect of the 
uncertainty in the parameter 
$\epsilon$, which controls the shape of the $\Delta\eta$ distribution
in Eq.~1, is $<1\%$. 

Our results for the ratio $R^{SDD}_{SD}$ 
of the number of events with a gap of $\Delta\eta>3$ 
to the total number of SD events are 
$0.246\pm 0.001\,{\rm (stat)}\pm 0.042\,{\rm (syst)}$ 
[$0.184\pm 0.001\,{\rm (stat)}\pm 0.043\,{\rm (syst)}$] at $\sqrt s=1800$ 
[630] GeV. These ratios are plotted in Fig.~4 at $\sqrt{s'}=$  
493 [173] GeV, the average value of the diffractive mass $M_X$, and 
compared with double diffractive to total cross section ratios 
$R^{DD}_{T}=\sigma^{DD}/\sigma_T$, where $\sigma^{DD}$ is obtained 
from~\cite{CDF_DD} and $\sigma_T$ is set to the Pomeron exchange 
contribution, $\beta^2(0)\left(\frac{s}{s_0}\right)^{\epsilon}$,
to conform with the definition of $R^{SDD}_{SD}$. 
The vertical error bars are mainly due 
to systematic effects, which are correlated among all points. 
The dashed lines represent 
predictions based on Regge theory and factorization normalized to 
the SD cross section at $\sqrt s=$22 GeV (see~\cite{multigap}). 
The solid lines are 
predictions from the ``renormalized gap probability" model~\cite{multigap},
in which the Regge cross section is factorized into two parts, 
one representing the $\bar pp$ total cross section at
the diffractive sub-energy    
multiplied by $\kappa^n$, where $n$ is the number of gaps, 
and a factor interpreted as 
the gap probability distribution 
normalized to unity over all available phase space.
The bands around the solid lines represent a 10\% uncertainty due to the 
factor $\kappa$~\cite{GM}. The data are in good agreement 
with the renormalized gap model 
predictions.

\begin{figure}
\centerline{\psfig{figure=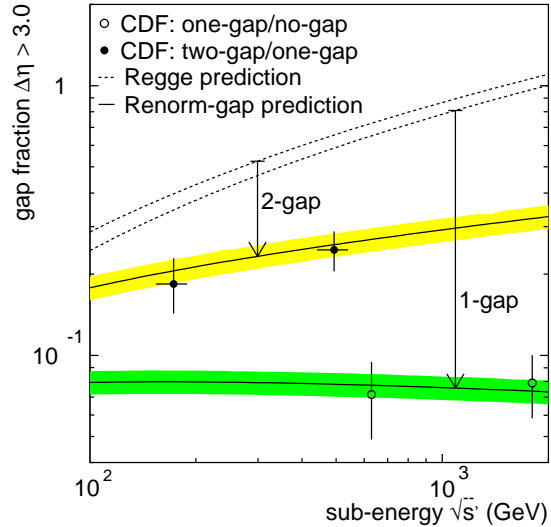,width=3in}}
\caption{Ratios of SDD (single-diffraction plus gap) to single-diffractive  
rates (filled circles)
and double-diffractive to total cross sections (open circles) as a function 
of the collision energy of the sub-process, Pomeron-proton and $\bar pp$, 
respectively. 
The uncertainties are mainly due 
to systematic effects, which are highly correlated among all four data points. 
The dashed lines are predictions from Regge theory and the solid lines 
from the renormalized gap probability 
model~\protect\cite{multigap}.}
\end{figure}
In summary, we have measured the fraction of events with a pseudorapidity gap 
$\Delta\eta>3$ within the diffractive cluster $X$ of the process 
$\bar pp\rightarrow \bar pX$ and found it to be 
$0.246\pm 0.001\,{\rm (stat)}\pm 0.042\,{\rm (syst)}$ 
[$0.184\pm 0.001\,{\rm (stat)}\pm 0.043\,{\rm (syst)}$] for 
$0.06<\xi_{\bar p}<0.09$ and $|t_{\bar p}|<1.0$ [0.2] GeV$^2$ at $\sqrt s=$1800
[630] GeV. These values are higher than expectations from double-diffractive
fractions in minimum bias events, lower than expectations from Regge theory and
factorization, and in good 
agreement with predictions based on the renormalized gap 
probability model~\cite{multigap}.  
 
We thank the Fermilab staff and the technical staffs of the
participating institutions for their vital contributions.  This work was
supported by the U.S. Department of Energy and National Science Foundation;
the Italian Istituto Nazionale di Fisica Nucleare; the Ministry of Education,
Culture, Sports, Science and Technology of Japan; 
the Natural Sciences and Engineering
Research Council of Canada; the National Science Council of the Republic of
China; the Swiss National Science Foundation; the A. P. Sloan Foundation; the
Bundesministerium fuer Bildung und Forschung, Germany; and the Korea Science
and Engineering Foundation.

\end{document}